# Density functional study of X monodoped and codoped (X=C, N, S, F) anatase $TiO_2$

Pengyu Dong[1]*, Huanhuan Pei[2], Qinfang Zhang[1] and Yuhua Wang[2]

[1]*Key Laboratory for Advanced Technology in Environmental Protection of Jiangsu Province, Yancheng Institute of Technology, Yancheng 224051, P. R. China*

[2]*Department of Materials Science, School of Physical Science and Technology, Lanzhou University, Lanzhou 730000, P. R. China*

*Corresponding author.

Tel.: +86-515-88298923 (Office), +86-18351484051(Mobile)

E-mail address: dongpy11@gmail.com

**Abstract**

Using density-functional theory (DFT) calculations within the generalized gradient corrected approximation, the models that nonmetallic impurities X (X=C, N, S, F) substituted for O or Ti sites in anatase $TiO_2$ were investigated. By calculating the oxygen vacancy formation energy of X-monodoped $TiO_2$ with X substituted for O, we suggested that X dopants existed as $C^{4-}$, $N^{3-}$, $S^{2-}$ and $F^{-}$ ions, respectively. Meanwhile, the X dopants existed as $C^{4+}$, $N^{3+}$, and $S^{6+}$ for X-monodoped (X=C, N, S) $TiO_2$ with X substituted for Ti. The conclusion of the valence states of nonmetallic impurities X substituted for O or Ti sites in $TiO_2$ is also supported by the results of optimized cell parameters and the local structures. Furthermore, an effective nonmetallic passivated codoping approach to modify the band edges of $TiO_2$ is proposed. Based on the first-principle calculations, we suggested that nonmetallic passivated groups such as ($S^{2-}$ + $C^{4+}$) and ($C^{4-}$ + $S^{6+}$) could reduce the band gap largely and make less perturbation in conduction band minima (CBM), thus lead to an ideal visible–light absorption region without affecting the reducing power.

*Keywords*: DFT; Oxygen vacancy; Valence state; Visible light.

## 1. Introduction

Among various semiconductors, $TiO_2$ has been widely employed to promote photocatalytic degradation of harmful organic compounds [1]. However, due to the wide band-gap, pure anatase $TiO_2$ photocatalyst mainly absorbs ultraviolet (UV) photons, which amounts for 4–5% of the incoming solar energy on the earth's surface. To utilize solar energy effectively, much effort has been directed toward the shift of the optical response of $TiO_2$ from UV to the visible spectral range. One approach is the doping of various transition metal [2, 3] and nonmetal elements [4, 5]. However, it is noted that the doping of transition metal elements could act as carrier recombination centers [2].

The pioneering work of nonmetal-doped $TiO_2$ was made by Asahi *et al.* in 2001, [6] who reported nitrogen-doped $TiO_2$ and suggested N 2p level could mix with O 2p states. The mixture states resulted in the narrowing of band gap and photocatalytic activity in visible light. Since then, nonmetallic anion doping has become the most attractive subject. Anionic nonmetal dopants such as nitrogen [7], carbon [8–10], sulfur [11], and fluorine [12] atoms have been widely investigated for the extension of photocatalytic activity into the visible-light region. However, the nature of the nonmetallic anion-induced modifications of $TiO_2$ electronic band structure is also controversial. Band gap narrowing [6, 13] or formation of localized midgap states [8, 14, 15] has been alternatively proposed. To obtain insight into this issue, we have performed a careful and systematic analysis of the nonmetallic anion impurity states as well as their influence on the $TiO_2$ band structure.

Compared to anion doping, some experimental works demonstrated that the nonmetal cation doped-TiO$_2$ also exhibited photocatalytic activity under visible light irradiation [16–19]. In particular, Ohno *et al.* synthesized chemically modified TiO$_2$ photocatalyst in which S (S$^{4+}$) and C (C$^{4+}$) substituted for some of the lattice Ti atoms, and they also found that both S$^{4+}$-doped and C$^{4+}$-doped TiO$_2$ showed excellent photocatalytic activity under visible light irradiation [17–19]. However, the valence state of S atom substituting for Ti atom in TiO$_2$ is uncertain (S$^{4+}$ or S$^{6+}$) in their works. Generally, the nonmetal substituting for a Ti site could lead to a significantly X–O bond, and the valence states would become complex and uncertain due to the similar electronegativity between O and X. Therefore, it is necessary to make certain of the valence states of nonmetal cation substituting for Ti atom in TiO$_2$.

In addition, Gai *et al.* proposed a passivated codoping approach to shift the TiO$_2$ absorption edge into visible light range and improve its photocatalytic efficiency for hydrogen production [20]. They suggested (C$^{4-}$ + Mo$^{6+}$)-codoped TiO$_2$ with both C anions and Mo cations could reduce the recombination centers and then enhance the photocatalytic activity. This is because the electrons on the donor levels could passivate the same amount of holes on the acceptor levels in the (C$^{4-}$ + Mo$^{6+}$)-codoped TiO$_2$ system. The next experimental works according to the above theoretical calculation were completed by Zhang *et al.* [21] and our group [22], respectively. However, their experimental results showed that the transition metal (Mo$^{6+}$) doped into TiO$_2$ often acts as the carrier recombination center and decreases the photocatalytic activity. Therefore, it seems that the most anticipated approach is the cooperation of anion and cation nonmetal-codoping, which would result in stronger visible light absorption and prevent the formation of recombination centers.

In this work, the band structures of nonmetallic impurities X-doped (X=C, N, S, F) anatase $TiO_2$ with X substituted for O or Ti sites are systematically investigated using DFT calculations. By calculating the oxygen vacancy formation energy of X-monodoped $TiO_2$, the valence states of X dopants are clarified. Moreover, a nonmetal passivated codoping approach to modify the band edges of $TiO_2$ is proposed.

## 2. Calculation method

The calculations performed in this work are based on the density-functional theory using the full-potential linear-augmented-plane-wave (FLAPW) method, while the generalized gradient approximation (GGA) is employed for the exchange-correlation potential. The scalar-relativistic effects are included for the band states, while the core-level states treated fully relativistically. The muffin-tin (MT) sphere radii RMT are chosen as 1.86 a.u. for Ti, and 1.65 a.u. for O, respectively. The cutoff parameter RMTKmax for limiting the number of the plane waves is 7.0 and the largest vector Gmax is 12.0. We simulated the X doping effects using $2 \times 2 \times 2$ (96-atom) repetition of the unit bulk anatase $TiO_2$ and the stable structure was assumed when the atomic force is less than 1 mRy/a.u.. For the Brillouin zone integration, we used 50 samping K points to calculate density of state (DOS) and partial density states (PDOS). The calculated band gap is 1.96 eV, much smaller than the experimental band gap of 3.20 eV. [23] The band gap underestimation of DFT calculations always exists due to the well-known limitation of predicting accurate excited states properties.

**3. Results and discussion**

3.1 Electronic structures and valence states in monodoped $TiO_2$

*3.1.1 The substitution for oxygen (anion monodoped $TiO_2$).*

The electronic structures of X-doped $TiO_2$ with X replaced an O site are examined. Fig. 1 shows the calculated DOS and PDOS for anion monodoped anatase $TiO_2$. The relative positions of DOS are adjusted by referencing to the core levels of the atom farthest from the impurity. The substitution of X on oxygen lattice site induces the isolated states in the band, which indicated that the anion dopants induced the formation of localized midgap states instead of band gap narrowing. The electronegativity of X increases in the order C < S < N < F (i.e., 2.48, 2.69, 2.90 and 3.91, respectively), while that of O is 3.41. The position of dopants' level induced by C, N, S and F decreases in the order C > S > N > F, which is corresponding to the electronegativity. Moreover, the highest localized states of C 2p, N 2p and S 3p orbitals in the PDOS are isolated inside the gap of $TiO_2$ (i.e., 1.05, 0.89, 0.14 eV above the VBM, respectively), which result in the photocatalytic activity in visible light region. The peak in the PDOS located at -8~-6 eV belongs to the 2p orbitals of F atoms is lower than O 2p states, which did not contribute to the reduction of the optical band gap. Furthermore, it should be noted that the C 2s and N 2s states are located at much lower energies (at about -8 and -12 eV, respectively), which is impossible to participate in bonding with O or Ti.

In principle, the substitution with an element of different charge can induce a charge imbalance, resulting in the formation of a crystallographic point defect such as oxygen or titanium vacancy. Recently, DFT calculation of N-doped TiO$_2$ [24] gave a successful explanation for the relative relationship between nitrogen and oxygen vacancy. The formation energies of oxygen vacancy in pure anatase (eqn. 1) and the doped TiO$_2$ (eqn. 2) were calculated in this work to find out the relationship and clarify the valence states of impurities, as shown in Table 1.

$$TiO_2 \rightarrow TiO_{2-x} + xV_o + 1/2xO_2 \qquad (1)$$

$$TiO_{2-y}X_y \rightarrow TiO_{2-x-y}X_y + xV_o + 1/2xO_2 \qquad (2)$$

For pure anatase TiO$_2$, the oxygen vacancy formation energy is 4.09 eV, while the oxygen vacancy formation cost is reduced in the presence of C and N doping (0.78 and 1.21 eV, respectively). This could be because the excess electrons associated to vacancies are trapped at dopant sites. On the contrary, the oxygen vacancy formation energy in S-doped TiO$_2$ is 4.18 eV, which is close to that of pure anatase TiO$_2$. Thus, we could conclude that S in the TiO$_2$ shows the same states of charge with O. The substitution of F for O (F$^-$ doping at an O$^{2-}$ site) could convert one Ti$^{4+}$ to Ti$^{3+}$, [25, 26] and the oxygen vacancy formation energy raises from 4.09 to 8.81 eV. These computational results suggest that the C, N, S and F replaced O$^{2-}$ ion exist as C$^{4-}$, N$^{3-}$, S$^{2-}$ and F$^-$ ion, respectively.

*3.1.2 The substitution for titanium (cation monodoped TiO$_2$).*

It was found that replacing a Ti site is favorable under O-rich growth condition from previous theory studies of nonmetal-doped TiO$_2$ [11, 27]. In order to make clear the valence states of the dopants in Ti site, the electronic structures of X-doped $TiO_2$ with X replaced a Ti atom were examined. The calculated DOS and PDOS are shown in Fig. 2. It can be seen that X-doped $TiO_2$ have mostly Ti 3d character from DOS when nonmetals are used to substitute for Ti site. From PDOS, it is observed that the C 2p, N 2p and S 3p states are contributed to conduction band, while the C 2s, N 2s and S 3s states are contributed to valence band. The N 2s state is located slightly above the valence band minima (VBM) with about 0.10 eV, while C 2s and S 3s are located much higher with about 0.78 and 0.85 eV, respectively. The electronic structure of F-doped $TiO_2$ is strikingly different from that of C-, N- and S-doped $TiO_2$. It is observed that F 2p states are contributed to valence band instead of conduction band. Moreover, the VBM of F-doped $TiO_2$ is less changed compared with that of pure $TiO_2$, which is less interesting for researchers.

The previous works indicated that the dopant atoms at Ti sites should exist as $C^{4+}$, $N^{5+}$, $S^{6+}$ ($s^2$ electrons are supplied for system) or $C^{2+}$, $N^{3+}$, $S^{4+}$ ($s^2$ electrons are restricted). Here, we assumed that C replaced Ti as $C^{2+}$, which must be accompanied by oxygen vacancy. This is because $C^{2+}$ ion has two less valence electrons than $Ti^{4+}$, and the doped $TiO_2$ is overall electrically neutral, thus the charge imbalance induces the formation of oxygen vacancy:

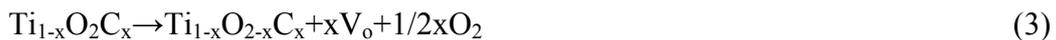

$Ti_{1-x}O_2C_x \rightarrow Ti_{1-x}O_{2-x}C_x + xV_o + 1/2xO_2$ (3)

However, the calculated oxygen vacancy formation energy for C-doped $TiO_2$ is 4.08 eV, which is close to that of pure $TiO_2$ (Table 2), indicating that C doping does not affect the charge balance. Therefore, C replaced Ti site should exist as $C^{4+}$ ion. On the contrary, the cost of the oxygen vacancy formation energy for N doping is 1.58 eV, suggesting the N dopant exists as $N^{3+}$. This is because $N^{3+}$ has one less valence electron than $Ti^{4+}$, which facilitates the formation of oxygen vacancy. We also checked the condition of S cation doping, and it is found that the oxygen vacancy formation energy rises from 4.09 to 7.27 eV, suggesting that the S dopant restricts the formation of oxygen vacancy and exists as $S^{6+}$ ion. In summary, the nonmetal X dopants (X=C, N, S) substituting for Ti site should exist as $C^{4+}$, $N^{4+}$ and $S^{6+}$, respectively. These conclusions are in contrast to the DFT calculation results by Yang *et al.* [11, 27], but are consistent with some experimental results [16, 18, 19].

3.2. Optimized structure

The effect of the X dopants on the structure of X-doped $TiO_2$ can be observed from the cell parameters and the local geometry around X dopants. X anion-doped $TiO_2$ with X replaced an O site induces the local $XTi_3$ structure, whereas for X cation substituting it is the local $XO_6$ structure (Fig. 3). The optimized unit cell parameters and the bond lengths are summarized in Table 3 and Table 4. It can be seen that the optimized structural parameters of pure $TiO_2$ are agreement with the experimental work (a=b=3.782Å, c=9.502Å, $d_{eq}$=1.932Å and $d_{ap}$=1.979Å). For X-doped $TiO_2$ with X replaced O, the optimized lattice parameters (a, b and c) are all greater than that of undoped $TiO_2$, and the a-axis length becoming greater than the b-axis length for each doping system (see Table 3). Moreover, the axial X−Ti ($X-Ti_{ap}$) bond lengths are much longer than the equatorial X−Ti ($X-Ti_{eq}$) bond lengths for each doping system. Since the $XTi_3$ structure has local C2v symmetry with one $X-Ti_{ap}$ bond as the rotational axis and two X–Ti$_{eq}$ bonds approximately along the a-direction, the a-axis length increases faster than b-axis when the radius of X becomes larger. [27] For X-doped TiO$_2$ with X replaced Ti, we also optimized the cell parameters using the supercell geometry. The optimized cell parameters (a, b and c) are greater than that of undoped TiO$_2$ (see Table 4), which is consistent with that of X anion-doped TiO$_2$. However, the bond lengths in the local XO$_6$ structure become complex: C–, N– and S–O bond lengths are smaller than the Ti–O bond length. Although the radius of F atom is smaller than C, N and S, the F–O bond length is larger than the Ti–O bond length. The most likely reason is the biggest electronegativity of F, restricting its electrons in participating in the bonding between F and O. In contrast, C 2p, N 2p and S 3p states show strong hybridization with O 2p orbitals in a bandwidth from -5~0 eV surrounding the valence band, as shown in Fig. 2. The bonding makes the atoms more closely associated than that of the undoped TiO$_2$ due to the smaller radius of dopants compared with Ti atom. The local XO$_6$ becomes almost symmetrical, and the axial N–O bond becomes shorter than the equatorial bond for N-doped TiO$_2$, which are contrary to those of pure anatase and C-doped TiO$_2$. The previous studies of nonmetal cation-doped TiO$_2$ [8, 11] suggested that the midgap band states are composed of the impurity and the O 2p states of its neighboring O atoms. However, we find that the midgap states consist of the dopant and two oxygen atoms in different sites only in XO$_6$ structure (O$_{eq}$–C–O$_{eq}$, O$_{eq}$–N–O$_{aq}$ and O$_{aq}$–S–O$_{aq}$, respectively) after rechecking the PDOS of all atoms in the local XO$_6$ octahedron structure. The reason for this change is the different valence states of impurities and the local XO$_6$ geometry: the excess electrons associated to the impurity are trapped by four O atoms in the equatorial plane firstly, and the axial two O sites are the second choice, which will lead to a shorter axial bond length. Therefore, when S occupied at Ti site, four 3d electrons are trapped at all of the equatorial O sites and the extra 3s electrons are trapped at the two axial O sites forming the mostly symmetrical octahedron (i.e., S–O$_{eq}$=1.890 Å, S–O$_{aq}$=1.886 Å). And one 2s electron of N is bonded with an axial O 2p orbital which still belongs to N atom inducing a shorter axial bond length. Whereas C$^{4+}$-doped and pure anatase TiO$_2$ have the same structure due to the same valence state with Ti$^{4+}$.

3.3. Passivated codoping in TiO$_2$

The above computational results suggest that the C, N and S atoms occupied at a Ti site exist as C$^{4+}$, N$^{3+}$ and S$^{6+}$ ions, respectively, and with the fact of C$^{4-}$, N$^{3-}$, S$^{2-}$ and F$^-$ ions at an O site. We proposed an approach to modify the band gap of TiO$_2$ with codoping group such as (C$^{4-}$ + S$^{6+}$), (S$^{2-}$ + C$^{4+}$), and (F$^-$ + N$^{3+}$). In these cases, the defect complex will not be an effective recombination center because the charge-compensated donor-acceptor pairs could passivate each other's electrical activities. Fig. 4 shows the DOS of (C$^{4-}$ + S$^{6+}$), (S$^{2-}$ + C$^{4+}$) and (F$^-$ + N$^{3+}$)-codoped TiO$_2$ systems compared with that of pure TiO$_2$. It can be seen that the changes in band gap caused by the donor-acceptor passivated codoping follow the same trends as that observed in the corresponding monodoped cases. For (F$^-$ + N$^{3+}$) codoping, there is a similar band structure with N$^{3+}$ monodoped TiO$_2$ because F-monodoping does not cause any change in the band edge of TiO$_2$. In contrast, the (C$^{4-}$ + S$^{6+}$) and (S$^{2-}$ +

$C^{4+}$)-codoped systems could have much stronger visible light absorption. This is because the highest midgap states are located 1.34 eV and 0.90 eV above the VBM, respectively, reducing the band gap.

The codoping-induced changes on the band gap are summarized in Fig. 5. The calculated band gap of pure $TiO_2$ is 1.96 eV, which is smaller than the actual experimental value 3.20 eV. Since the experimental band gap of undoped $TiO_2$ could be obtained by adding 1.24 eV to the calculated band gap, the underestimation of band gaps was corrected by adding 1.24 eV to the calculated value. From Fig. 5, it can be seen that both ($C^{4-}$ + $S^{6+}$) and ($S^{2-}$ + $C^{4+}$) codoping are the proper systems. This is because they reduce the band gap largely and less perturbation in CBM, which are greatly ideal for visible light response and do not effect the reducing power. To examine if these defect pairs are stable, the binding energy of defect pair ($E_b$) was calculated by following formula [28]:

$$E_b = E_{tot}(X_O + Y_{Ti}) + E_{tot}(TiO_2) - E_{tot}(X_O) - E_{tot}(Y_{Ti}) \qquad (4)$$

where $E_{tot}$ is the total energy calculated using the same supercell. The binding energies $E_b$ for the ($C^{4-}$ + $S^{6+}$), ($S^{2-}$ + $C^{4+}$) and ($F^-$ + $N^{3+}$) pairs are -3.22, -2.79 and 1.17 eV, respectively. The negative $E_b$ indicates that the pairs are stable due to the formation of defect complex in which $Y_{Ti}$ donates its electrons to $X_O$, resulting in large interaction between anion acceptor and cation donor. For ($F^-$ + $N^{3+}$) codoping system, the charges transfer from the neighboring $Ti^{4+}$ to $F^-$, resulting in the positive $E_b$ with the unstable structure.

**4. Conclusions**

The DOS of X-doped $TiO_2$ with X replaced an O site indicated that the anion dopants induced the formation of localized midgap states instead of band gap narrowing. For nonmetal anion-doped $TiO_2$, the X dopants existed as $C^{4-}$, $N^{3-}$, $S^{2-}$ and $F^-$ ions, respectively. On the contrary, the valence states of impurities become complex for nonmetal cation-doped $TiO_2$ due to the similar electronegativity of X dopants with O. The changes of bond lengths in local $XO_6$ within the cation-doped $TiO_2$ are associated with the valence states of X dopants. In addition, we suggest that both ($C^{4-}$+ $S^{6+}$) and ($S^{2-}$ + $C^{4+}$) codoping methods are the proper systems for visible-light photocatalysis based on the first-principles calculations.

**Acknowledgements**

Research is financially supported by the National Science Foundation for Distinguished Young Scholars (No. 50925206).

**Figure captions:**

Fig. 1 DOS (A) and PDOS (B) for 96-atom anatase supercell with X replaced O: (a) C-doped, (b) N-doped, (c) S-doped, and (d) F-doped. The energy is measured from the top of the valence band of $TiO_2$.

Fig. 2 DOS (A) and PDOS (B) for 96-atom anatase supercell with X replaced Ti: (a) C-doped, (b) N-doped, (c) S-doped, and (d) F-doped. The energy is measured from the top of the valence band of $TiO_2$.

Fig. 3 Schematic models for (a) local structure $OTi_3$ with X substituted for O, (b) local structure $TiO_6$ with X substituted for Ti.

Fig. 4 The calculated DOS of passivated codoping systems (red) compared with pure $TiO_2$ (black).

Fig. 5 The band structures of passivated codoping anatase $TiO_2$ from the FLAPW after modified by the experimental value of undoped $TiO_2$. The hydrogen production level and the water oxidation level are displayed with dashed lines.

Table 1 Oxygen vacancy formation energy of pure anatase and X anion-doped anatase TiO$_2$.

|  | Pure | C&O | N&O | S&O | F&O |
|---|---|---|---|---|---|
| *(eV) | 4.09 | 0.78 | 1.21 | 4.18 | 8.81 |

Table 2 Oxygen vacancy formation energy of pure anatase and X cation-doped anatase TiO$_2$.

|   | Pure | C&Ti | N&Ti | S&Ti |
|---|---|---|---|---|
| *(eV) | 4.09 | 4.08 | 1.58 | 10.27 |

Table 3 Optimized structural parameters of undoped and X-doped anatase $TiO_2$ (X = N, C, S, F) with X replaced O.

|  | Undoped | C&O | N&O | S&O | F&O |
|---|---|---|---|---|---|
| a(Å) | 3.768 | 3.843 | 3.829 | 3.887 | 3.815 |
| b(Å) | 3.768 | 3.809 | 3.813 | 3.827 | 3.808 |
| c(Å) | 9.458 | 9.563 | 9.552 | 9.485 | 9.546 |
| X–Ti$_{eq}$(Å) | 1.921 | 1.966 | 1.942 | 2.002 | 1.922 |
| X–Ti$_{ap}$(Å) | 1.988 | 2.192 | 2.024 | 2.232 | 2.002 |

Table 4 Optimized structural parameters of undoped and X-doped anatase TiO$_2$ (X = N, C, S, F) with X replaced Ti.

|  | Undoped | C&Ti | N&Ti | S&Ti | F&Ti |
| --- | --- | --- | --- | --- | --- |
| a(Å) | 3.768 | 3.810 | 3.803 | 3.837 | 3.795 |
| b(Å) | 3.768 | 3.809 | 3.804 | 3.837 | 3.796 |
| c(Å) | 9.458 | 9.563 | 9.552 | 9.485 | 9.546 |
| X–O$_{eq}$(Å) | 1.921 | 1.827 | 1.874 | 1.890 | 1.949 |
| X–O$_{ap}$(Å) | 1.988 | 1.876 | 1.802 | 1.886 | 2.415 |

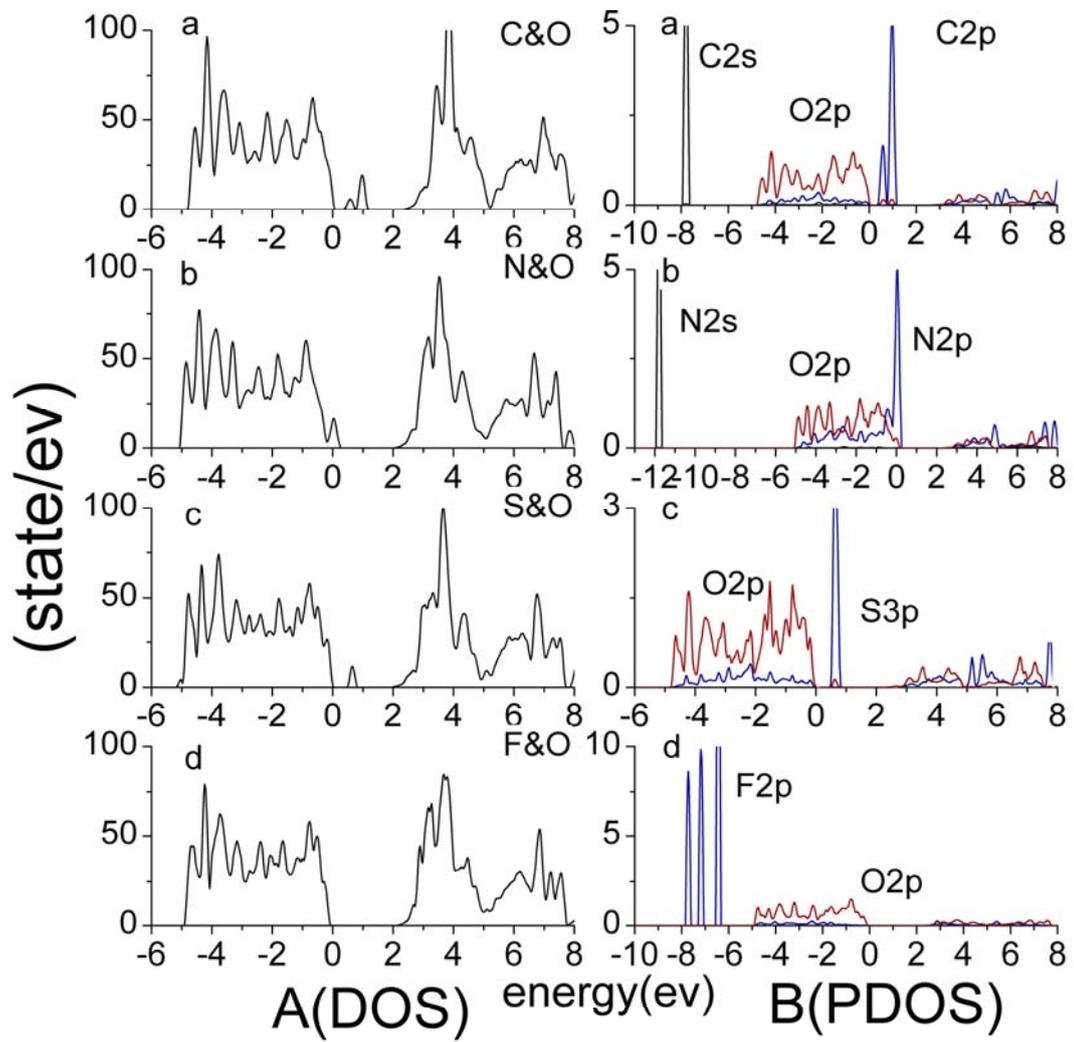

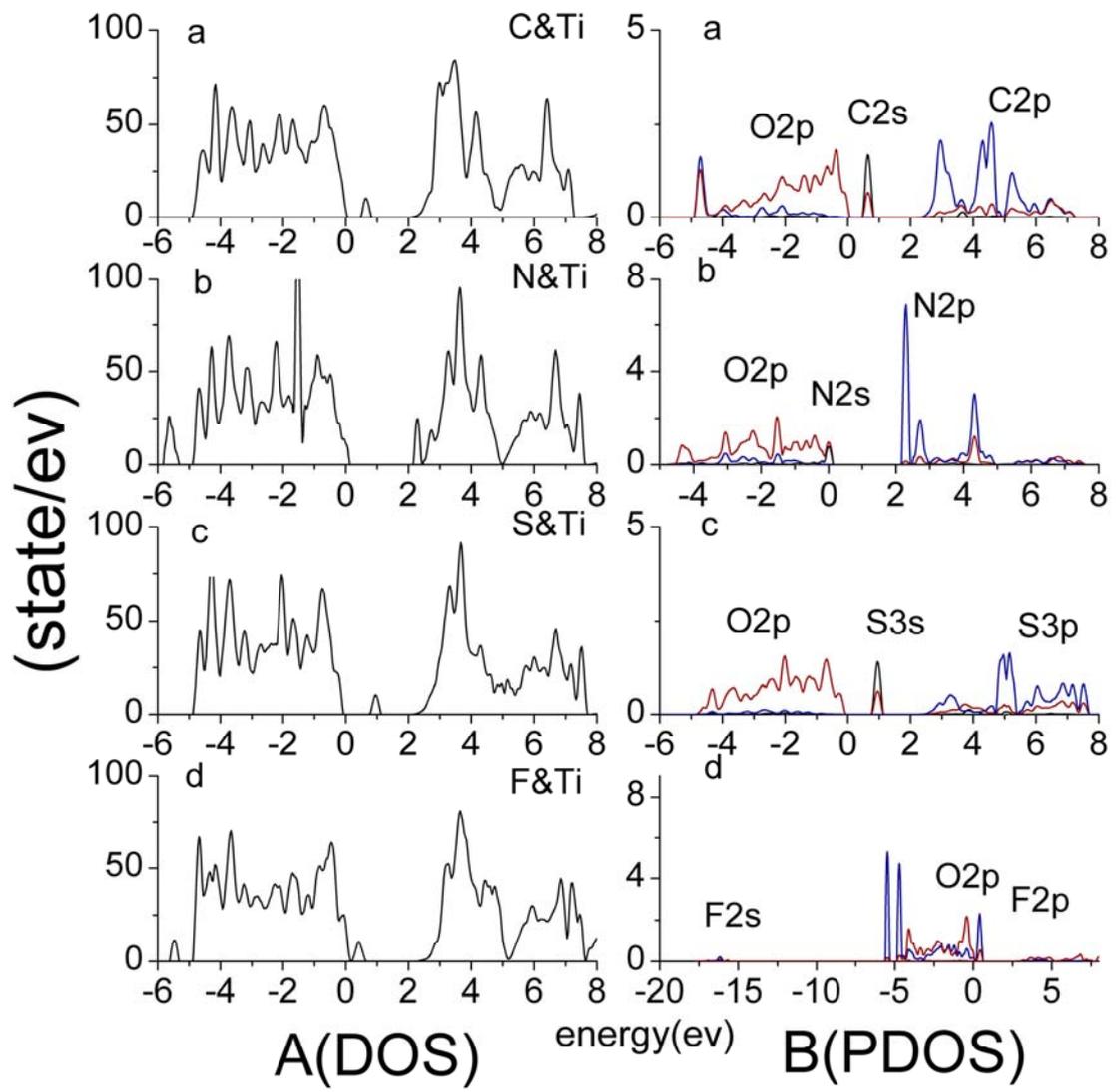

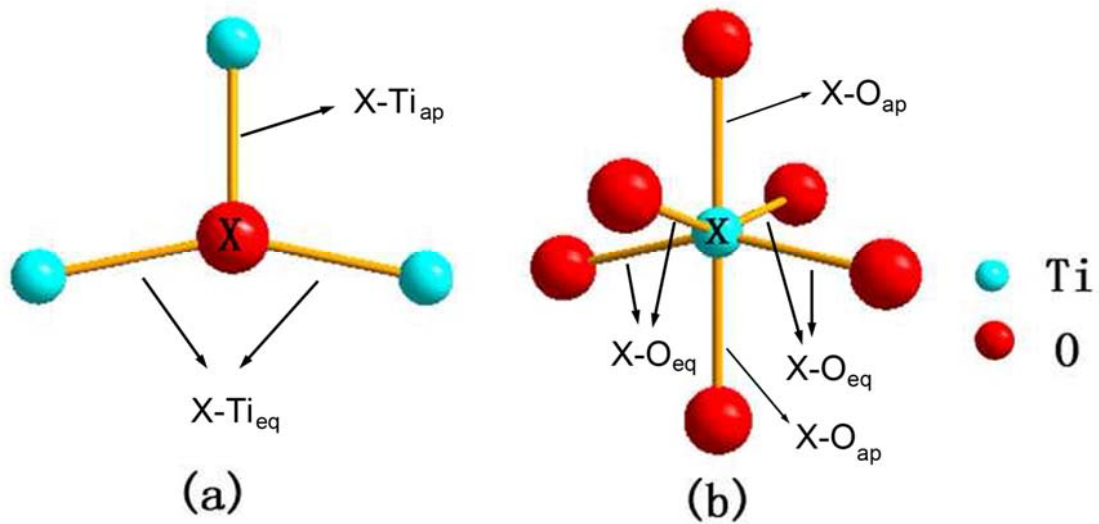

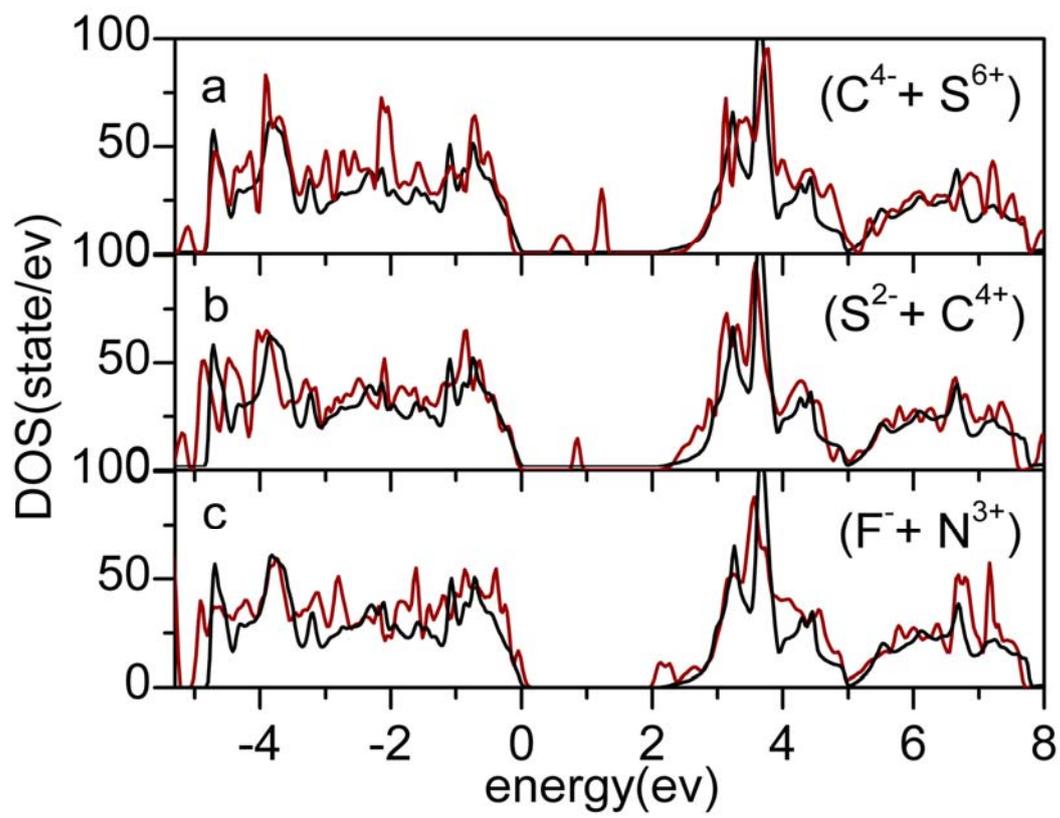

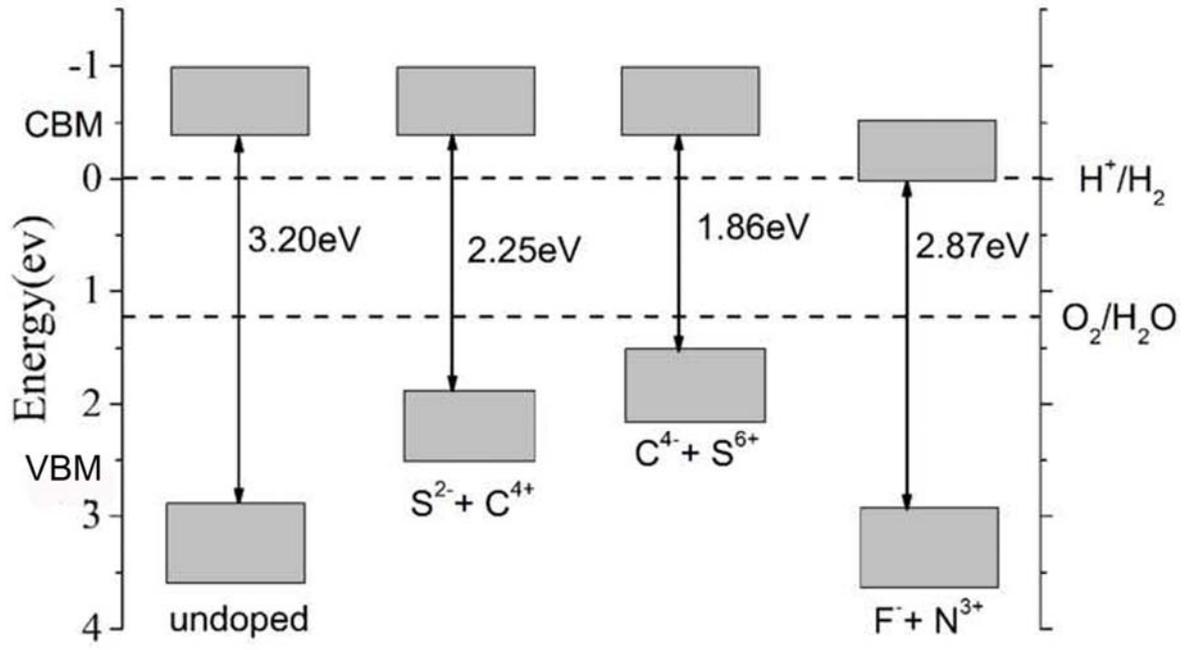25